\documentclass[journal,twocolumn]{IEEEtran}
\usepackage{url}
\usepackage{cmap}
\usepackage[T1]{fontenc}
\usepackage{amsmath,graphicx,hhline}
\usepackage{caption}
\usepackage{subcaption}
\usepackage{amssymb,amsthm}
\usepackage{mathrsfs}
\usepackage[nodisplayskipstretch]{setspace}
\usepackage{enumerate}
\usepackage{multirow}
\usepackage[update,prepend]{epstopdf}
\usepackage{float}
\restylefloat{table}
\usepackage{listings}
\usepackage{algorithm}
\usepackage{algpseudocode}
\usepackage{booktabs}
\usepackage{lipsum}
\usepackage{geometry}
\usepackage{paralist}
\usepackage{cite}

\newgeometry{
    top=1in,
    bottom=1in,
    outer=1in,
    inner=1in,
}

\newtheorem*{theorem*}{Theorem}
\newtheorem{theorem}{Theorem}
\newtheorem*{problemstatement*}{Problem Statement}
\newtheorem*{definition*}{Definition}

\newtheorem*{Lemma*}{Lemma}
\newtheorem*{corollary*}{Corollary}

\newtheorem{property}[theorem]{Property}

\title{Spectral Projector-Based Graph Fourier Transforms}
\author{Joya A. Deri,~\IEEEmembership{Member, IEEE,}
        and Jos\'{e} M. F. Moura,~\IEEEmembership{Fellow, IEEE}
        \thanks{This work was partially supported by NSF grants CCF-1011903 and CCF-1513936 and an SYS-CMU grant.}
        \thanks{The authors are with the Department of Electrical and Computer Engineering, Carnegie Mellon University, Pittsburgh, PA 15213 USA (email: {jderi,moura}@andrew.cmu.edu)}%
        }

\begin{document}
\maketitle
\begin{abstract}
 The paper presents the graph Fourier transform (GFT) of a signal in terms of its spectral decomposition over the Jordan subspaces of the graph adjacency matrix~$A$. This representation is unique and coordinate free, and it leads to unambiguous definition of the spectral components (``harmonics'') of a graph signal. This is particularly meaningful when~$A$ has repeated eigenvalues, and it is very useful when~$A$ is \emph{defective} or not diagonalizable (as it may be the case with directed graphs). Many real world large sparse graphs have defective adjacency matrices. We present properties of the GFT and show it to satisfy a generalized Parseval inequality and to admit a total variation ordering of the spectral components. We express the GFT in terms of spectral projectors and present an illustrative example for a real world large urban traffic dataset.
\end{abstract}
\section{Introduction}
\label{sec:intro}
Graph signal processing~(GSP) extends traditional signal processing to data indexed by nodes of graphs. Such data arises in many domains from genomics to business to social networks, to name a few. In GSP, the graph Fourier transform~(GFT) has been defined through the eigendecomposition of the adjacency matrix~$A$ of the graph, taken as the graph shift operator~\cite{sandryhaila2013discrete,sandryhaila2014big,sandryhaila2014discrete}, or of the graph Laplacian~$L$ \cite{shuman2013emerging}. In the GSP approach in~\cite{sandryhaila2013discrete,sandryhaila2014big,sandryhaila2014discrete} and according to the algebraic signal processing in~\cite{puschel2008algebraic_foundation,puschel2008algebraic_1dspace,puschel2008algebraic_cooleytukey} the eigenvectors of the shift are the \textit{graph frequency} or \textit{graph spectral components} and the eigenvalues are the \textit{graph frequencies}.

\textbf{Contributions}. There are several issues that need further study:
\begin{inparaenum}[1)]
    \item \textit{Unicity}: the matrix form of the GFT in~\cite{sandryhaila2013discrete,sandryhaila2014big,sandryhaila2014discrete,shuman2013emerging} is not unique, depending on (implicit or explicit) choice of bases for underlying spaces. This is true, even if the matrix of interest is diagonalizable;
    \item \textit{Spectral components}: If~$A$ or~$L$ have repeated eigenvalues, there may be several eigenvectors corresponding to this repeated eigenvalue (frequency)\textemdash defining the spectral or frequency components (``harmonics'') becomes an issue; and
    \item \textit{Nondiagonalizability}: If the shift is not diagonalizable, as it happens in many real world applications with large sparse graphs, the matrix~$A$ is defective, introducing additional degrees of freedom in the coordinate form definition of the GFT.
\end{inparaenum}
These topics are particularly relevant when applying GSP to datasets arising in real world problems. In many of these, the graphs are large and sparse and their adjacency matrix is defective.

This paper addresses these issues. We present the coordinate free GFT that leads to a unique spectral decomposition of graph signals and to an unambiguous definition of spectral components, regardless if there are repeated eigenvalues or not, or if the shift is defective. Spectral components \cite{puschel2008algebraic_foundation} are signals that are left invariant by graph filters. For repeated eigenvalues or defective matrices, it makes sense to consider in this context \textit{irreducible invariant} subspaces\textemdash signal subspaces that are invariant to filtering and are irreducible. This is achieved by decomposing the signal space into a direct sum of irreducible filter-invariant (spectral) subspaces. If the dimension of the filter-invariant subspaces is larger than one, the choice of basis for these subspaces is not unique, and neither is the coordinate form of the GFT or identifying spectral components with basis vectors. The coordinate free GFT and the spectral decomposition we present successfully addresses these challenges. We present a spectral oblique projector-based GFT that allows for a  unique and  unambiguous spectral representation of a signal over defective adjacency matrices. Invariance to filtering follows from invariance to the shift operator (adjacency matrix~$A$) since, by GSP~\cite{sandryhaila2013discrete,sandryhaila2014big,sandryhaila2014discrete}, shift invariant filters are polynomials in the shift~$A$. The spectral components are the Jordan subspaces of the adjacency matrix. We show that the GFT allows characterization of the signal projection energies via a generalized Parseval's identity.
%
Total variation ordering of the spectral components with respect to the Jordan subspaces is also discussed.

\textbf{Synopsis of approach.} Before we formally introduce the concepts and as a way of introduction and motivation, we explain very concisely our approach. From algebraic signal processing (ASP)~\cite{puschel2008algebraic_foundation,puschel2008algebraic_1dspace}, we know that the basic component is the \emph{signal processing model} $\Omega=(\mathcal A, \mathcal M, \Phi)$. For a vector space~$V$ of complex-valued signals, we can then generalize for this  signal model~$\Omega$, linear filtering theory, where algebra~$\mathcal A$ corresponds to a filter space, module~$\mathcal M$  corresponds to a signal space, and bijective linear mapping~$\Phi: V \rightarrow \mathcal M$ generalizes the~$z$-transform~\cite{puschel2008algebraic_foundation}. One way to create a signal model is to specify a generator (or generators) for~$\mathcal A$, the shift filter or shift operator. The Fourier transform is the map from the signal module~$\mathcal M$ to an irreducible decomposition of~$\mathcal M$ where the irreducible components are invariant to the shift (and to the filters). We are then interested in studying the invariant irreducible components of~$\mathcal M$. These are the Jordan subspaces as explained below. In GSP, we choose as shift the adjacency matrix~$A$ of the underlying graph. Similarly, then, the Jordan subspaces play an important role in the graph Fourier transform defined in Section~\ref{sec:gft} and, in this context, the irreducible, $\mathcal A$-invariant submodules~$\mathcal M'\subseteq \mathcal M$ are the \emph{spectral components} of (signal space)~$\mathcal M$. The Jordan subspaces are invariant, irreducible subspaces of $\mathbb C^N$ with respect to the adjacency matrix~$A$; they  represent the spectral components. This motivates the definition of a spectral projector-based graph Fourier transform in Section~\ref{sec:gft}.

Section~\ref{sec:prevwork} describes related spectral analysis methods and graph signal processing frameworks. Section~\ref{sec:background} provides the graph signal processing and linear algebra background for the graph Fourier transform defined in Section~\ref{sec:gft}. Section~\ref{sec:gft:parseval} presents the generalized Parseval's identity as a method for ranking spectral components. Total variation-based orderings of the Jordan subspaces are discussed in detail in Section~\ref{sec:jordanequiv:totalvar}. Section~\ref{sec:application} shiows an application on a real world dataset. Limitations of the method are briefly discussed in Section~\ref{sec:limitations}.
\section{Previous work}
This section presents a brief review of the literature and some background material.
\label{sec:prevwork}
%
%
%
\subsection{Spectral methods}
\label{subsec:spectralmethods}
Principal component analysis (the Karhunen-Lo\`{e}ve Transform) is an early signal decomposition method  proposed and remains a fundamental tool today. This approach orthogonally transforms data points, often via eigendecomposition or singular value decomposition (SVD) of an empirical covariance matrix,  into linearly uncorrelated variables called principal components~\cite{pearson1901pca,hotelling1933analysis,hotelling1936relations}. The first principal components capture the most variance in the data; this allows analysis to be restricted to these first few principal components, thus increasing the efficiency of the data representation. 

Other methods determine low-dimensional representations of high-dimensional data by projecting the data onto low-dimensional subspaces generated by subsets of an eigenbasis~\cite{tenenbaum2000,roweisSaul2000,belkinNiyogi2003,donohoGrimes2003}. References~\cite{tenenbaum2000,roweisSaul2000} embed high-dimensional vectors onto low-dimensional manifolds determined by a weight matrix with entries corresponding to nearest-neighbor distances. In~\cite{belkinNiyogi2003}, embedding data in a low-dimensional space is described in terms of the graph Laplacian, where the graph Laplacian is an approximation to the Laplace-Beltrami operator on manifolds. 
Reference~\cite{belkinnyogi-2008} also proves that the algorithm~\cite{tenenbaum2000} approximates eigenfunctions of a Laplacian-based matrix.

These methods~\cite{tenenbaum2000,roweisSaul2000,belkinNiyogi2003,donohoGrimes2003} focus on discovering low-dimensional representations for high-dimensional data, capturing relationships between data variables into a matrix for their analysis. In contrast, our problem treats the data as a signal that is an input to a graph-based filter. Our approach emphasizes node-based weights (the signal) instead of edge-based weights that capture data dependencies.   Related node-based methods in the graph signal processing framework are discussed next.


\textbf{Data indexed by graphs and Laplacian-based GFTs.}
The graph signal processing framework developed in this paper assumes that data is indexed by graphs.  Studies that analyze data indexed by nodes of a graph include~\cite{ganesanstorage2005,wagner2005distributed,wagner2006architecture}, which use wavelet transforms to study data on distributed sensor networks. Other approaches, such as those in~\cite{coifmanMaggioni2006,shuman2013emerging,hammond2011wavelets,narang2012perfect,agaskarinftheory2012,ekambaram2013,zhu2012approximating}, use the graph Laplacian and its eigenbasis for localized data processing. In particular,~\cite{hammond2011wavelets,shuman2013emerging} define a graph Fourier transform (GFT) as signal projections onto the Laplacian eigenvectors. These eigenvectors form an orthonormal basis since the graph Laplacian is symmetric and positive semidefinite. Graph-based filter banks are constructed with respect to this GFT in~\cite{narang2012perfect}.

Analyses based on the graph Laplacian do not take into account first-order network structure of the network, that is, any asymmetry like in a digraph or directed edges in a graph. These asymmetries affect network flows, random walks, and other graph properties, as studied, for example, in~\cite{chung2016digraphrandomwalks,chung2012spanningtreediameter}. The approach we take here preserves the influence of directed edges in graph signal processing by projecting onto the eigenbasis of the adjacency matrix.

\textbf{Adjacency matrix-based GFTs.} References~\cite{sandryhaila2013discrete,sandryhaila2014big,sandryhaila2014discrete} develop GSP, including filtering, convolution, graph Fourier transform,  from the graph adjacency matrix~$A\in\mathbb C^{N\times N}$, taken to play the role of shift operator~$z^{-1}$ in digital signal processing. According to the algebraic signal processing theory of~\cite{puschelSIAM2003,puschel2008algebraic_foundation,puschel2008algebraic_1dspace,puschel2008algebraic_cooleytukey}, 
the shift generates  all linear shift-invariant filters for a class of signals (under certain shift invariance assumptions). In the context of GSP~\cite{sandryhaila2013discrete}, shift invariant filters are polynomials on the shift~$A$. The graph Fourier transform is defined also in terms of the adjacency matrix. GSP as presented in~\cite{sandryhaila2013discrete,sandryhaila2014big,sandryhaila2014discrete} preserves the directed network structure, in contrast to second order methods like those based on the graph Laplacian.

The graph Fourier transform of~\cite{sandryhaila2013discrete} is defined as follows. For a graph~$\mathcal G=G(A)$ with adjacency matrix $A\in\mathbb C^{N\times N}$ and Jordan decomposition $A=VJV^{-1}$,  the graph Fourier transform of a signal $s\in\mathbb C^{N}$ over~$\mathcal G$ is defined as
\begin{equation}
\label{eq:origgft}
\widetilde s = V^{-1}s,
\end{equation}
where $V^{-1}$ is the \emph{Fourier transform matrix} of $\mathcal G$. This is essentially a projection of the signal onto the eigenvectors of $A$. It is an orthogonal projection when $A$ is normal ($A^H A= A A^H$) and the eigenvectors form a unitary basis (i.e., $V^{-1}= V^H$). This is of course guaranteed with the graph Laplacian. Left unanswered in these approaches is the lack of unicity\footnote{Eigenvalues are defined up to a constant. Different choices lead to scaled polynomial transforms. The discrete Fourier transform corresponds to a very specific choice of basis~\cite{puschel2008algebraic_foundation}.} of $V^{-1}$, the appropriate definition of spectral components when there are repeated eigenvalues, and finally how to define it uniquely when the shift is defective.

This paper addresses these topics and, in particular, focuses on graph signal processing over \emph{defective}, or non-diagonalizable, adjacency matrices. These matrices have at least one eigenvalue with algebraic multiplicity (the exponent in the characteristic polynomial of $A$) greater than the geometric multiplicity (the kernel dimension of $A$), which results in an eigenvector matrix that does not span $\mathbb C^N$.

The basis can be completed by computing Jordan chains of generalized eigenvectors~\cite{lancaster1985,golub2013matrix}, but the computation introduces degrees of freedom that render these generalized eigenvectors non-unique; in other words, the transform~\eqref{eq:origgft} may vary greatly depending on the particular generalized eigenvectors that are chosen. Our approach defines the GFT in terms of spectral projections onto the Jordan subspaces (i.e., the span of the Jordan chains) of the adjacency matrix\footnote{We recognize that computing the Jordan decomposition is numerically unstable. This paper is focused on the concepts of a spectral projection coordinate free definition of the GFT and spectral components. Section~\ref{sec:limitations} will address these computational issues that, for lack of space, are fully discussed in~\cite{deriAIM2016,deriEquiv2016}.}.


%
%
%
%
\section{Background}
\label{sec:background}
This section reviews the concepts of graph signal processing and provides a reference for the underlying mathematics. Section~\ref{sec:background:gsp} defines the graph Fourier transform and graph filters; see also~\cite{sandryhaila2013discrete,sandryhaila2014big,sandryhaila2014discrete}. Section~\ref{sec:background:eigendecomp} defines the generalized eigenspaces and Jordan subspaces of a matrix~\cite{lancaster1985,golub2013matrix,gohberg2006invariant}.

\begin{figure}[t]
\centerline{\includegraphics[width=.6\linewidth]{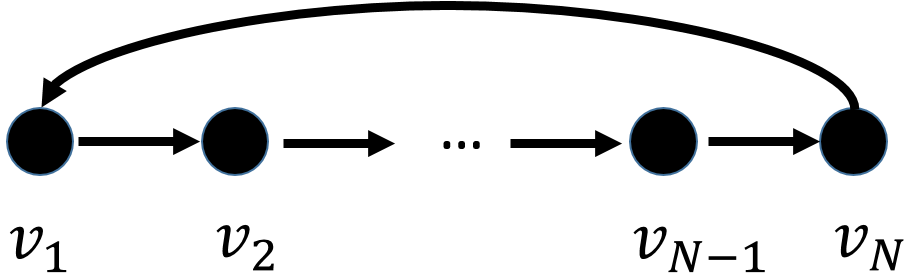}}
\caption{Directed cycle graph.}
\label{fig:cyclegraph}
\end{figure}
\subsection{Graph Signal Processing}
\label{sec:background:gsp}
\subsubsection{Graph Signals}
\label{sec:background:gsp:sigs}
Let~$\mathcal G = \mathcal G(A)= (\mathcal V,\mathcal E)$ be the graph corresponding to matrix~$A\in\mathbb C^{N\times N}$, where $\mathcal V$ is the set of~$N$ nodes and a nonzero entry~$\left[A\right]_{ij}$ denotes a directed edge $e_{ij}\in\mathcal E$ from node~$j$ to node~$i$.  In real-world applications, such nodes can be represented by geo-locations of a road network, and the edges can be specified by one-way or two-way streets. Define \emph{graph signal}~$s: \mathcal V \rightarrow \mathcal S$ on $\mathcal G$, where~$\mathcal S$ represents the signal space over the nodes of $\mathcal G$.  We take~$\mathcal S = \mathbb C^N$ such that $s=(s_1,\dots,s_N)\in\mathbb C^N$ and~$s_i$ represents a measure at node $v_i\in\mathcal V$.
In real-world applications, such signals can be specified by sensor measurements or datasets.
\begin{figure*}[tb]
\centering
\includegraphics[width=.6\linewidth]{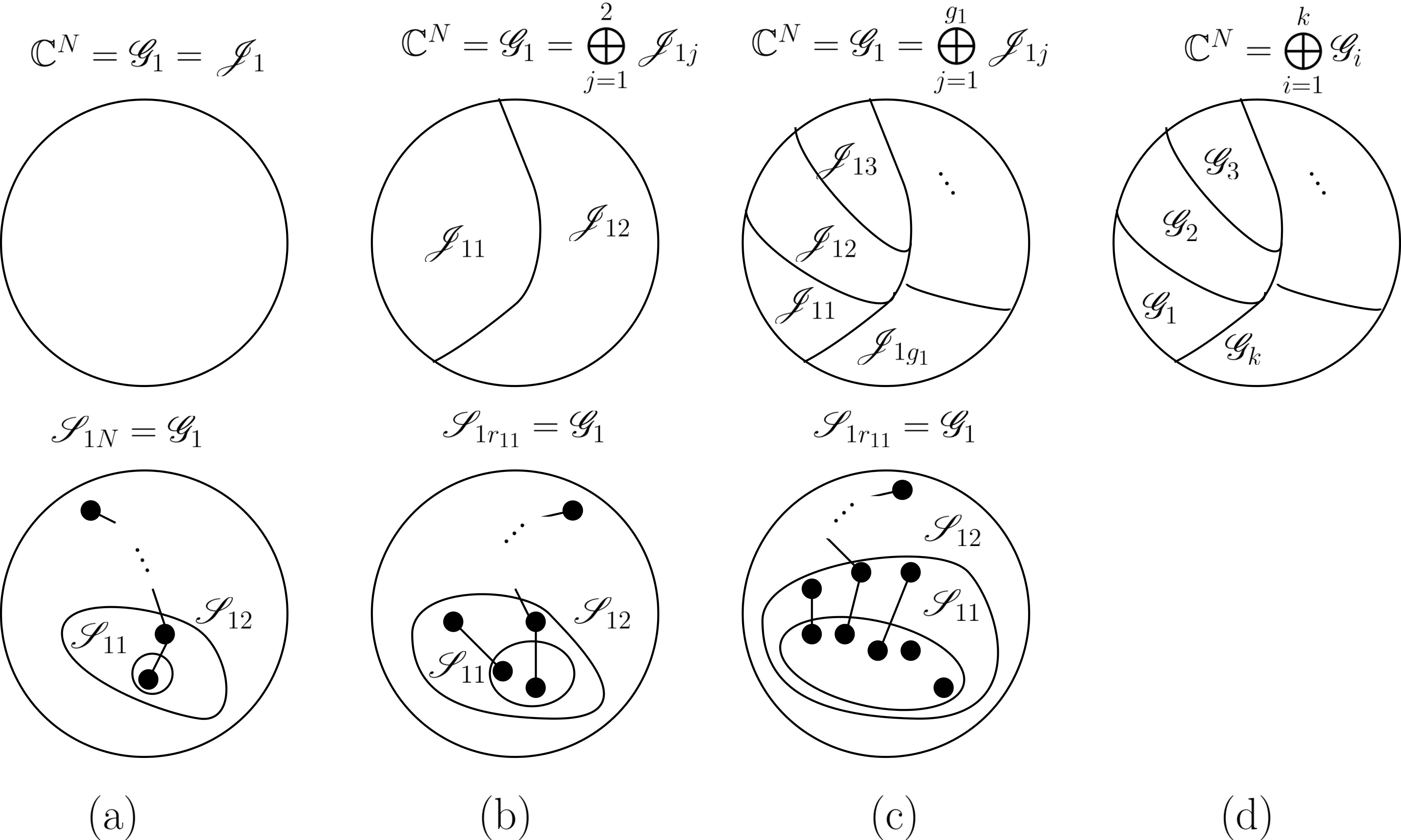}
\caption{Illustration of generalized eigenspace partitions and Jordan chains of adjacency matrix~$A\in\mathbb C^{N\times N}$ for (a) a single Jordan block, (b) one eigenvalue and two Jordan blocks, (c) one eigenvalue and multiple Jordan blocks, and (d) multiple eigenvalues. In (a)-(c) (bottom), each point represents a vector in a Jordan chain of~$A$; points connected by lines illustrate a single Jordan chain. The partial multiplicities depicted for~$\lambda_1$ are (a)~$N$, (b)~$r_{11} = N-2$ and~$2$, and (c)~$r_{11} = N-6$, $2$, $2$, $1$, and $1$. 
Each generalized eigenspace $\mathscr G_i$ in (d) can be visualized by (a)-(c).}
\label{fig:jordan_explain}
\vspace{-.10cm}
\end{figure*}

\subsubsection{Graph Shift}
\label{sec:background:gsp:adjmat}
As in~\cite{sandryhaila2013discrete,sandryhaila2014discrete}, the graph shift is the graph signal processing counterpart to the shift operator~$z^{-1}$ in digital signal processing.  The graph shift is defined as the operator that replaces the element~$s_i$ of graph signal~$s=(s_1,\dots,s_N)$ corresponding to node $v_i\in V$ with the linear combination of the signal elements at its in-neighbors (nodes $v_k\in V$ that participate in an edge $e_{ik}\in\mathcal E$), denoted by set~$\mathcal N_i$; i.e., the shifted signal has elements~$\widetilde s_i = \sum_{v_j\in\mathcal N_i} \left[A\right]_{ij} s_j$, or
\begin{equation}
\label{eq:graphshift}
\widetilde s = A s.
\end{equation}
Consistency with discrete signal processing can be seen by considering the directed cycle graph in Figure~\ref{fig:cyclegraph}, which represents a finite, periodic time-series signal. The adjacency matrix of the graph is circulant matrix (elements not shown are zero)
\begin{equation}
\label{eq:circulantmatrix}
C = \begin{bmatrix}
  &&&&& 1 \\
1     \\
  && \ddots\\
&   &&& 1
\end{bmatrix}.
\end{equation}
The shift $\widetilde s = C s$ yields the time delay~$\widetilde s_i = s_{i-1\,\, \mathrm{mod}\, N}$.

Reference~\cite{sandryhaila2013discrete} shows that the graph shift motivates defining the graph Fourier transform as the signal projection onto the eigenvectors of~$A$. Our transform in Section~\ref{sec:gft} builds on this concept to develop a framework to handle defective adjacency matrices.
\subsubsection{Graph Filter}
\label{sec:background:gsp:filters}
The graph shift is a simple graph filter, where a graph filter~$\mathbf H\in \mathbb C^{N\times N}$ represents a (linear) system with output~$\mathbf Hs$ for any graph signal~$s\in\mathcal S$.  As shown in Theorem~1 of~\cite{sandryhaila2013discrete}, graph filter~$\mathbf H$ is shift-invariant, or
\begin{equation}
\label{eq:shiftinvar}
A\left(\mathbf H\right)s = \mathbf H \left( As\right),
\end{equation}
if and only if a polynomial $h(x) = \sum_{i=0}^{L} h_i x^i$ exists for constants $h_0,h_1,\dots,h_L\in\mathbb C$ such that $\mathbf H = h(A) = \sum_{i=0}^{L} h_i A^i$. This condition holds whenever the characteristic and minimal polynomials of~$A$ are equal~\cite{sandryhaila2013discrete}. 

For defective~$A$ with unequal characteristic and minimal polynomials such as the examples seen in this paper, shift-invariance cannot be claimed; however, an \emph{equivalent graph filter} can be designed in terms of a matrix that is the image of a polynomial of~$A$~\cite{sandryhaila2013discrete}. The properties of such graph filters are established in~\cite{sandryhaila2013discrete}.

\subsection{Eigendecomposition}
\label{sec:background:eigendecomp}
%
This section and Appendix~\ref{app:JordanDecomp} provide a review of Jordan decompositions. The reader is directed to~\cite{lancaster1985,golub2013matrix,gohberg2006invariant,horn2012matrix} for additional background. 
Jordan subspaces and the Jordan decomposition are defined in this section.  

The generalized eigenspaces $\mathscr G_i = \mathrm{Ker}(A-\lambda_i I)^{m_i}$  of $A\in\mathbb C^{N\times N}$ corresponding to its~$k$ distinct eigenvalues~$\lambda_i$ decompose $\mathbb C^N$ in terms of the direct sum
\begin{equation}
\label{eq:genspaces_decomp}
\mathbb C^N = \bigoplus_{i=1}^{k} \mathscr G_i
\end{equation}
as depicted in Figure~\ref{fig:jordan_explain}d; see also Appendix~\ref{app:JordanDecomp}.

Let $v_{1} \in\mathrm{Ker}(A-\lambda_i I)$, $v_1\neq 0$, be one of the $g_i$ proper eigenvectors of~$A$ corresponding to the eigenvalue~$\lambda_i$. It generates by recursion  the generalized eigenvectors
\begin{equation}
\label{eq:jordanrecurse_subspace}
Av_{p} = \lambda_i v_{p} + v_{p-1}, \:\: p=2,\dots,r
\end{equation}
where~$r$ is the minimal positive integer such that $\left(A-\lambda_i I\right)^r v_r=0$ and $\left(A-\lambda_i I\right)^{r-1} v_r\neq0$. 
  Such a sequence of vectors $(v_{1},\dots,v_{r})$ that satisfies~\eqref{eq:jordanrecurse_subspace} is a \emph{Jordan chain of length~$r$}. The Jordan chain vectors are linearly independent and span a  \emph{Jordan subspace}, or,
\begin{equation}
\label{eq:jordansubspace}
\mathscr J = \mathrm{span}\left(v_1,v_2,\dots,v_r\right).
\end{equation}

Order the Jordan subspaces of $\lambda_i$ by decreasing dimension and denote by $\mathscr J_{ij}$ the $j$th Jordan subspace of $\lambda_i$ with dimension $r_{ij}\leq m_i$, where $\{r_{ij}\}_{j=1}^{g_i}$ are called the \emph{partial multiplicities} of $\lambda_i$.  Then the generalized eigenspace~$\mathscr G_i=\mathrm{Ker} (A-\lambda I)^{m_i}$ of $\lambda_i$ can be decomposed as
\begin{equation}
\label{eq:jordansubspace_decomp}
\mathscr G_i = \bigoplus_{j = 1}^{g_i}\mathscr J_{ij}
\end{equation}
as depicted in Figures~\ref{fig:jordan_explain}a-c. Combining~\eqref{eq:jordansubspace_decomp} and~\eqref{eq:genspaces_decomp}, the Jordan subspaces uniquely decompose~$\mathbb C^N$ as
\begin{equation}
\label{eq:CN-jordansubspace_decomp}
\mathbb C^N = \bigoplus_{i = 1}^k\bigoplus_{j = 1}^{g_i}\mathscr J_{ij}.
\end{equation}
Furthermore, the cyclic Jordan subspaces cannot be represented as direct sums of smaller invariant subspaces; that is, the Jordan subspaces are \emph{irreducible} components of $\mathbb C^N$ (see, e.g., p.~318 of~\cite{gohberg2006invariant}). 

Figure~\ref{fig:jordan_explain}  illustrates possible Jordan subspace structures of~$A$, with the top row showing the tessellation of  vector space~$\mathbb C^N$ by the generalized or root eigenspace $\mathscr G_i=\mathrm{Ker} \left(A-\lambda_i I\right)^{m_i}$ and by the Jordan spaces $\mathscr J_{ij}$, and the bottom row illustrating the telescoping of~$\mathbb C^N$ by the generalized eigenspaces of order~$p$.
Figure~\ref{fig:jordan_explain}a illustrates $\mathbb C^N$ for a matrix with a single Jordan chain, represented by connected points in $\mathbb C^N$.
The case of a matrix with two Jordan blocks corresponding to one eigenvalue is shown in Figure~\ref{fig:jordan_explain}b. 
Figure~\ref{fig:jordan_explain}c shows $\mathbb C^N$ for a matrix with a single eigenvalue and multiple Jordan blocks, and Figure~\ref{fig:jordan_explain}d depicts the tessellation of the space in terms of the generalized eigenspaces for the case of multiple distinct eigenvalues.

\textbf{Jordan decomposition.} Appendix~\ref{app:JordanDecomp}  defines eigenvector matrix $V$ and Jordan normal form $J$ such that $A=VJV^{-1}$. An important property of this decomposition is that Jordan chains are not unique and not necessarily orthogonal. For example, the $3\times 3$ matrix
\begin{equation}
\label{eq:exmatrix}
\setlength\arraycolsep{12pt}
A = \begin{bmatrix}
0 & 1 & 1 \\ 0 & 0 & 1\\ 0 & 0 & 0
\end{bmatrix}
\end{equation}
can have distinct eigenvector matrices
\begin{equation}
\label{eq:exmatrix_V}
V_1 = \left[\begin{array}{lrr}
1 & -1 & 1 \\ 0 & 1 & -2\\ 0 & 0 & 1
\end{array}\right],
V_2 = \left[\begin{array}{lrr}
1 & 0 & 0 \\ 0 & 1 & -1\\ 0 & 0 & 1
\end{array}\right],
\end{equation}
where the Jordan chain vectors are the columns of~$V_1$ and~$V_2$ and so satisfy~\eqref{eq:jordanrecurse_subspace}.  Since Jordan chains are not unique, the Jordan subspace is used in Section~\ref{sec:gft} to characterize the possible generalized eigenvectors. 
%
%
%
%
%
\section{Spectral Projector-Based Graph Signal Processing}
\label{sec:gft}
This section presents a basis-invariant coordinate free graph Fourier transform with respect to a set of known proper eigenvectors. For graphs with diagonalizable adjacency matrices, the coordinate form of this transform resolves with appropriate choice of basis vectors to that of~\cite{sandryhaila2013discrete,sandryhaila2014discrete}. The interpretation of the spectral components is settled for the cases of repeated eigenvalues and non-diagonalizable, or \emph{defective}, adjacency matrices.

Consider matrix~$A$ with distinct eigenvalues~$\lambda_1,\dots,\lambda_k$, $k\leq N$, that has Jordan decomposition~$A=VJV^{-1}$. Denote by~$\mathscr J_{ij}$ the~$j$th Jordan subspace of dimension~$r_{ij}$ corresponding to eigenvalue~$\lambda_i$,~$i=1,\dots,k$,~$j=1,\dots,g_i$. 
Each~$\mathscr J_{ij}$ is~$A$-invariant and irreducible (see Section~\ref{sec:background:eigendecomp}). Then,
  the Jordan subspaces are the spectral components of the signal space~$\mathcal S = \mathbb C^N$ 
and define the graph Fourier transform of a graph signal~$s\in\mathcal S$ as the mapping
\begin{align}
\mathcal F: \mathcal S &\rightarrow  \bigoplus_{i=1}^k \bigoplus_{j=1}^{g_i} \mathscr J_{ij}\nonumber\\
s&\rightarrow \left(\widehat s_{11},\dots,\widehat s_{1g_1},\dots,\widehat s_{k1},\dots,\widehat s_{kg_k}\right) \label{eq:gft}.
\end{align}
That is, the Fourier transform of~$s$, is the unique decomposition
\begin{equation}
\label{eq:gft_sum}
s = \sum_{i=1}^k\sum_{j=1}^{g_i} \widehat s_{ij},\hspace{1cm}\widehat s_{ij}\in \mathscr J_{ij}.
\end{equation}

The distinct eigenvalues $ \lambda_1,\dots,\lambda_k$ are the \emph{graph frequencies} of graph~$\mathcal G(A)$. The \emph{frequency} or \emph{spectral components} of graph frequency~$\lambda_i$ are the Jordan subspaces~$\mathscr {J}_{ij}$. The total number of frequency components corresponding to $\lambda_i$ is its geometric multiplicity $g_i$. In this way, when $g_i>1$, frequency~$\lambda_i$ corresponds to more than one frequency component.

To highlight the significance of~\eqref{eq:gft} and~\eqref{eq:gft_sum}, consider the signal expansion of a graph signal $s$ with respect to graph $\mathcal G(A)$:
\begin{equation}
\label{eq:gft_signalexpansion}
s = \widetilde s_1 v_1 +\cdots + \widetilde s_N v_N = V \widetilde s,
\end{equation}
where $v_i$ is the $i$th basis vector in a Jordan basis of~$A$, $V$ is the corresponding eigenvector matrix, and $\widetilde s_i$ is the $i$th expansion coefficient. As discussed in Section~\ref{sec:background:eigendecomp}, the choice of Jordan basis has degrees of freedom when the dimension of a cyclic Jordan subspace is greater than one. Therefore, if $\mathrm{dim} \mathscr J_{ij} \geq 2$, there exists eigenvector submatrix $X_{ij}\neq V_{ij}$ such that $\mathrm{span}\{X_{ij}\} = \mathrm{span}\{V_{ij}\} = \mathscr J_{ij}$. Thus, the signal expansion~\eqref{eq:gft_signalexpansion} is not unique when $A$ has Jordan subspaces of dimension~$r>1$.

In contrast, the Fourier transform given by~\eqref{eq:gft} and~\eqref{eq:gft_sum}  yields a unique signal expansion that is independent of the choice of Jordan basis. Given any Jordan basis $v_1,\dots,v_N$ with respect to $A$, the $j$th spectral component of eigenvalue~$\lambda_i$ is, by~\eqref{eq:gft_sum},~$\widehat s_{ij} = \sum_{k=p}^{p+r_{ij}-1} \widetilde s_kv_k$, where $v_p,\dots,v_{p+r_{ij}-1}$ are a basis of~$\mathscr J_{ij}$.  Under this definition, there is no ambiguity in the  interpretation of frequency components even when Jordan subspaces have dimension~$r>1$ or there are repeated eigenvalues.  The properties of the spectral components are discussed in more detail below.
\subsection{Spectral Components}
\label{sec:gft:def:spectralcomponents}
The spectral components of the Fourier transform~\eqref{eq:gft} are expressed in terms of basis $v_1,\dots,v_N$ and its dual basis $w_1,\dots,w_N$ since the chosen Jordan basis may not be orthogonal. Denote the basis and dual basis matrices by $V = [v_1 \cdots v_N]$ and $W = [w_1 \cdots, w_N]$. By definition,~$\langle w_i,v_j\rangle = w_i^H v_j= \delta_{ij}$, where $\delta_{ij}$ is the Kronecker delta function~\cite{horn2012matrix,jelena2014foundations}. Then~$W^H V = V^H W =I$, so the dual basis is the inverse Hermitian $W = V^{-H}$ and correspond to the left eigenbasis.

Partition Jordan basis matrix~$V$ as~\eqref{eq:V} so that each $V_{ij}\in\mathbb C^{N\times r_{ij}}$ spans Jordan subspace~$\mathscr J_{ij}$. Similarly, partition the dual basis matrix by rows 
as~$W = [\cdots W_{i1}^H\cdots  W_{ig_i}^H \cdots]^T$, with each~$W_{ij}^H\in\mathbb C^{r_{ij}\times N}$. Suppose $V_{ij}= [v_1 \cdots v_{r_{ij}}]$ with corresponding coefficients $\widetilde s_1,\dots, \widetilde s_{r_{ij}}$ in the Jordan basis expansion~\eqref{eq:gft_signalexpansion}. Define an~$N\times N$ matrix~$V^0_{ij} = [ 0 \cdots V_{ij} \cdots 0]$ that is zero except for the columns corresponding to~$V_{ij}$.
Then each spectral component corresponding to Jordan subspace~$\mathscr J_{ij}$ can be written as (below, diagonal dots and elements not shown are zero)
\begin{align}
\widehat s_{ij} &= \widetilde s_1 v_1 + \dots +\widetilde s_{r_{ij}} v_{r_{ij}} \label{eq:gft_singlecomponent_1}\\
&= V^0_{ij} \widetilde s \\
&=  V_{ij}^0 V^{-1} s \\
&= V \begin{bmatrix}
\ddots \\& I_{r_{ij}} \\ && \ddots
\end{bmatrix} V^{-1}s\\
&=   V \begin{bmatrix}
\ddots \\& I_{r_{ij}} \\ && \ddots
\end{bmatrix} W^{H} s \\
&= V_{ij} W^H_{ij} s,
\label{eq:gft_singlecomponent}
\end{align}
for  $i=1,\dots, k$ and $j=1,\dots,g_i$. Denote
\begin{equation}
\label{eq:spectralprojectormatrix}
P_{ij} = V_{ij} W^H_{ij},
\end{equation}
which is the \emph{projection matrix} onto $\mathcal S_{ij}$  parallel to complementary subspace~$\mathcal S\setminus \mathcal S_{ij}$. Note that~$P_{ij}^2 = V_{ij} W^H_{ij}V_{ij} W^H_{ij} = V_{ij} W^H_{ij}=P_{ij}$.

The projection matrices $\{P_{ij}\}_{j=1}^{r_{ij}}$ are related to the \emph{first component matrix}~$Z_{i0}$ of eigenvalue~$\lambda_i$. The component matrix is defined as~\cite[Section 9.5]{lancaster1985}
\begin{equation}
\label{eq:componentmatrix}
Z_{i0}  = V \begin{bmatrix}
\ddots \\& I_{a_i} \\ && \ddots
\end{bmatrix} V^{-1}
\end{equation}
where $a_i=\sum_{j=1}^{g_i} r_{ij}$ is the algebraic multiplicity of~$\lambda_i$. This matrix acts as a projection matrix onto the generalized eigenspace.

Theorem~\ref{thm:Pij} provides additional properties of projection matrix~$P_{ij}$.
\begin{theorem}
\label{thm:Pij}
For matrix~$A\in\mathbb C^{N\times N}$ with eigenvalues~$\lambda_1,\dots,\lambda_k$, the projection matrices $P_{ij}$ onto the~$j$th Jordan subspace~$\mathscr J_{ij}$ corresponding to eigenvalue~$\lambda_i$,~$i=1,\dots,k$, $j=1,\dots,g_i$, satisfy the following properties:
\begin{enumerate}[(a)]
\item $P_{ij}P_{kl} = \delta_{ik}\delta_{jl} P_{ij}$, where $\delta$ is the Kronecker delta function;
\item $\sum_{j=1}^{g_i} P_{ij} = Z_{i0}$, where~$Z_{i0}$ is the \emph{component matrix} of eigenvalue~$\lambda_i$;
\end{enumerate}
\end{theorem}
\begin{IEEEproof}
(a) Since $W^H V = I$, the partition of~$W^H$ and~$V$ that yields~\eqref{eq:gft_singlecomponent} satisfies~$W^H_{ij} V_{kl} = \delta_{ik}\delta_{jl} I_{r_{ij}\times r_{kl}}$, where $r_{ij}$ is the dimension of the Jordan subspace corresponding to~$P_{ij}$,~$r_{kl}$ the dimension of Jordan subspace corresponding to~$P_{kl}$, and matrix~$I_{r_{ij}\times r_{kl}}$ consists of the first~$r_{kl}$ canonical vectors~$e_i = (0,\dots,1,\dots,0)$, where~$1$ is at the~$i$th index. Then it follows that
\begin{align}
P_{ij}P_{kl} &= V_{ij} W^H_{ij}V_{kl} W^H_{kl} \\
&= V_{ij}\left(\delta_{ik}\delta_{jl} I_{r_{ij}\times r_{kl}}\right) W^H_{kl}.
\end{align}
If $i=k$ and $j=l$, then $P_{ij}P_{kl} = V_{ij} I_{r_{ij}\times r_{ij}} W^H_{ij} = P_{ij}$; otherwise, $P_{ij}P_{kl} = 0$.

(b) Write
\begin{align}
\sum_{j=1}^{g_i} P_{ij}  & = \sum_{j=1}^{g_i} V \begin{bmatrix}
\ddots \\& I_{r_{ij}} \\ && \ddots
\end{bmatrix} V^{-1}\\
&= V \left(\sum_{j=1}^{g_i} \begin{bmatrix}
\ddots \\& I_{r_{ij}} \\ && \ddots
\end{bmatrix}\right) V^{-1}\\
&= V \begin{bmatrix}
\ddots\\
& I_{\sum\limits_{j=1}^{g_i} r_{ij}}\\
&&\ddots
\end{bmatrix} V^{-1}\\
&= Z_{i0},
\end{align}
or the first component matrix of~$A$ for eigenvalue~$\lambda_i$. 
\end{IEEEproof}
Theorem~\ref{thm:Pij}(a) shows that each projection matrix~$P_{ij}$ only projects onto Jordan subspace~$\mathscr J_{ij}$. Theorem~\ref{thm:Pij}(b) shows that the sum of projection matrices for a given eigenvalue equals the component matrix of that eigenvalue.

This section provides the mathematical foundation for a graph Fourier transform based on projections onto the Jordan subspace of an adjacency matrix. The next section motivates ranking signal projections on Jordan subspaces by the energy of the signal projections.
\section{Generalized Parseval's Identity}
\label{sec:gft:parseval}
As discussed above, a chosen Jordan basis for matrix~$A\in\mathbb C^{N\times N}$, represented by the eigenvector matrix~$V$, may not be orthogonal. Therefore, Parseval's identity may not hold. Nevertheless, a generalized Parseval's identity does exist in terms of the Jordan basis and its dual; see also~\cite{jelena2014foundations}. For a dual basis matrix~$W = V^{-H}$, the following property holds:
\begin{property}[Generalized Parseval's Identity]
\label{prop:generalizedparseval}
Consider graph signals~$s_1,s_2\in\mathbb C^N$ over graph $\mathcal G(A)$,~$A\in\mathbb C^{N\times N}$. Let $V = [v_1 \cdots v_N]$ be a Jordan basis for~$A$ with dual basis $W = V^{-H}$ partitioned as $[w_1 \cdots w_N]$. Let $s = \sum_{i=1}^N \langle s,v_i \rangle v_i = V\widetilde s_V$ be the representation of $s$ in basis~$V$ and $s = \sum_{i=1}^N \langle s,w_i \rangle w_i = W\widetilde s_W$ be the representation of~$s$ in basis~$W$. Then
\begin{equation}
\label{eq:parseval}
\langle s_1,s_2  \rangle = \langle \widetilde s_{1,V},\widetilde s_{2,W} \rangle.
\end{equation}
By extension,
\begin{equation}
\label{eq:parseval_equalargs}
\left\| s\right\|^2 = \langle s,s  \rangle = \langle \widetilde s_{V},\widetilde s_{W} \rangle.
\end{equation}
\end{property}
Equations~\eqref{eq:parseval} and~\eqref{eq:parseval_equalargs} hold regardless of the choice of eigenvector basis.

\textbf{Energy of spectral components.}  The \emph{energy} of a discrete signal~$s\in\mathbb C^N$ is defined as~\cite{oppenheim1983signals,oppenheim1999}
\begin{equation}
\label{eq:energydef}
E_s = \langle s,s\rangle = \left\| s\right\|^2 = \sum_{i=1}^{N} \left| s\right|^2.
\end{equation}
Equation~\eqref{eq:parseval_equalargs} thus illustrates conservation of signal energy in terms of both a Jordan basis and its dual.

The energy of the signal projections onto the spectral components of~$\mathcal G(A)$ for the GFT~\eqref{eq:gft} is next defined. Write~$\widehat s_{i,j}$ in terms of the columns of~$V$ as $\widehat s_{i,j} = \alpha_1 v_{i,j,1}+\dots \alpha_{r_{ij}}v_{i,j,r_{ij}}$ and in terms of the columns of $W$ as  $\widehat s_{i,j} = \beta_1 w_{i,j,1}+\dots \beta_{r_{ij}}w_{i,j,r_{ij}}$.  Then the energy of~$\widehat s_{ij}$ can be defined as
\begin{equation}
\label{eq:gftenergy}
\|\widehat s_{ij}\|^2 = \langle \alpha,\beta\rangle
\end{equation}
using the notation $\alpha = (\alpha_1,\dots,\alpha_{r_{ij}})$ and $\beta = (\beta_1,\dots,\beta_{r_{ij}})$.


The generalized Parseval's identity expresses the energy of the signal in terms of the signal expansion coefficients $\widetilde s$, which highlights the importance of choosing a Jordan basis.  This emphasizes that both the GFT $\{\widehat s_{ij}\}$ and the signal expansion coefficients $\widetilde s$ are necessary to fully characterize the graph Fourier domain.


\textbf{Normal $A$.} When $A$ is normal (i.e., when $AA^H = A^H A$), $V$ can be chosen to have unitary columns. Then, $V = W$ so
\begin{equation}
\label{eq:parseval_normal}
\langle s_1,s_2  \rangle = \langle \widetilde s_{1},\widetilde s_{2} \rangle
\end{equation}
and
\begin{equation}
\label{eq:parseval_equalargs_normal}
\left\| s\right\|^2 = \langle s,s  \rangle = \left\| \widetilde s \right\|^2.
\end{equation}
Note that~\eqref{eq:parseval_normal} and~\eqref{eq:parseval_equalargs_normal} do not hold in general for diagonalizable~$A$.

While the Jordan basis, or choice of eigenvectors, is not unique, the image of a signal~$s$ under the projection matrix~$P_{ij}$~\eqref{eq:spectralprojectormatrix} is invariant to the choice of Jordan basis. This section shows that these projections can be ranked in terms of the percentage of recovered signal energy.

The next section demonstrates a total variation-based ranking of the spectral components.
%
%
%
%
%
\section{Total Variation Ordering}
\label{sec:jordanequiv:totalvar}
This section defines a mapping of spectral components to the real line to achieve an ordering of the spectral components. This ordering can be used to distinguish generalized low and high frequencies as in~\cite{sandryhaila2014discrete}. An upper bound for a total-variation based mapping of a spectral component (Jordan subspace) is derived.

As in~\cite{sandryhaila2014discrete,mallat2008}, the \emph{total variation} for finite discrete-valued (periodic) time series~$s$ is defined as
\begin{equation}
\label{eq:TV_signal}
\mathrm{TV}\left(s\right) = \sum_{n=0}^{N-1}  \left| s_n - s_{n-1\,\, \mathrm{mod} N}\right| = \left\|s - Cs\right\|_1,
\end{equation}
where $C$ is the circulant matrix~\eqref{eq:circulantmatrix} that represents the DSP shift operator. As in~\cite{sandryhaila2014discrete},~\eqref{eq:TV_signal} is generalized to the graph shift~$A$ to define the \emph{graph total variation}
\begin{equation}
\label{eq:TV_signal_graph}
\mathrm{TV}_G\left(s\right) =  \left\|s - A s\right\|_1.
\end{equation}
Matrix $A$ can be replaced by $A^{\mathrm{norm}} = \frac{1}{\left|\lambda_{\mathrm{max}}\right|} A$ when the maximum eigenvalue satisfies $|\lambda_{\mathrm{max}}|>0$.

Equation~\eqref{eq:TV_signal_graph} can be applied to define the {total variation} of a spectral component. These components are the cyclic Jordan subspaces of the graph shift~$A$ as described in Section~\ref{sec:background:eigendecomp}. 
Choose a Jordan basis of~$A$ so that~$V$ is the eigenvector matrix of~$A$, i.e., $A= VJV^{-1}$, where $J$ is the Jordan form of~$A$. Partition $V$ into $N\times r_{ij}$ submatrices $V_{ij}$ whose columns are a Jordan chain of (and thus span) the $j$th Jordan subspace~$\mathscr J_{ij}$ of eigenvalue~$\lambda_i$, $i=1,\dots, k\leq N$, $j = 1,\dots,g_{i}$.
Define the (graph) total variation of $V_{ij}$ as
\begin{equation}
\label{eq:TV_singlecomp}
\mathrm{TV}_G\left(V_{ij}\right) =  \left\|V_{ij} - A V_{ij} \right\|_1,
\end{equation}
where $\|\cdot\|_1$ represents the induced L1 matrix norm (equal to the maximum absolute column sum). 

The next theorem shows equivalent formulations for the graph total variation~\eqref{eq:TV_singlecomp}.
\begin{theorem}
\label{thm:jordanequiv_TVreformulated}
The graph total variation with respect to GFT~\eqref{eq:gft} can be written as
\begin{align}
&\mathrm{TV}_G\left(V_{ij}\right) = \left\|V_{ij}\left(I_{r_{ij}} - J_{ij} \right) \right\|_1\label{eq:TV_beforeineq}\\
&= \max_{i=2,\dots,r_{ij}} \left\{\left|1-\lambda\right| \left\|v_1\right\|_1, \left\|\left(1-\lambda\right) v_i - v_{i-1}\right\|_1\right\}.\label{eq:TV_jordanchainvecs}
\end{align}
\end{theorem}
\begin{IEEEproof}
Simplify~\eqref{eq:TV_singlecomp} to obtain
\begin{align}
\mathrm{TV}_G\left(V_{ij}\right) &=  \left\|V_{ij} - VJV^{-1} V_{ij} \right\|_1\\
&= \left\|V_{ij} - VJ \begin{bmatrix} \underline{0} \\ I_{r_{ij}} \\ \underline{0} \end{bmatrix} \right\|_1\\
&= \left\|V_{ij} - V \begin{bmatrix} \setlength\arraycolsep{20pt}
\underline{0} \\ & J_{ij} \\ &&\underline{0} \end{bmatrix} \right\|_1\\
&= \left\|V_{ij} - V_{ij}J_{ij}\right\|_1\\
&= \left\|V_{ij}\left(I_{r_{ij}} - J_{ij} \right) \right\|_1. 
\end{align}
Let $\lambda$ denote the $i$th eigenvalue and the columns $v_1,\dots,v_{r_{ij}}$ of $V_{ij}$ comprise its $j$th Jordan chain. Then~\eqref{eq:TV_beforeineq} can be expressed in terms of the Jordan chain:
\begin{align}
&\mathrm{TV}_G\left(V_{ij}\right)\\ &= \left\|\begin{bmatrix} v_1  \dots v_{r_{ij}}\end{bmatrix}
\begin{bmatrix} 1-\lambda & -1 \\ & 1-\lambda & \ddots\\ && \ddots &-1\\ &&&1-\lambda\end{bmatrix}\right\|_1\\
& = \left\|\begin{bmatrix} \left(1 - \lambda \right)v_1  & \left(1 - \lambda \right)v_2 - v_1 & \cdots & \end{bmatrix}\right\|_1\\
&= \max_{i=2,\dots,r_{ij}} \left\{\left|1-\lambda\right| \left\|v_1\right\|_1, \left\|\left(1-\lambda\right) v_i - v_{i-1}\right\|_1\right\}. 
\end{align}
\end{IEEEproof}
Theorem~\ref{thm:norm1Jordanchain} shows that~$V$ can be chosen such that $\left\|V_{ij}\right\|_1 = 1$ without loss of generality.
\begin{theorem}
\label{thm:norm1Jordanchain}
The eigenvector matrix~$V$ of adjacency matrix~$A\in\mathbb{C}^{N\times N}$ can be chosen so that each Jordan chain represented by the eigenvector submatrix $V_{ij}\in\mathbb C^{N\times r_{ij}}$ satisfies $\left\|V_{ij}\right\|_1 = 1$; i.e., $\left\|V \right\|_1 = 1$ without loss of generality.
\end{theorem}
\begin{IEEEproof}
Let $V$ represent an eigenvector matrix of $A$ with partitions $V_{ij}$ as described above, and let $J_{ij}$ represent the corresponding Jordan block. Let $D$ be a block diagonal matrix with $r_{ij}\times r_{ij}$ diagonal blocks $D_{ij}= (1/ \left\| V_{ij}\right\|_1) I_{r_{ij}}$.  Since~$D_{ij}$ commutes with $J_{ij}$, $D$ commutes with $J$. Note that $D$ is a special case of upper triangular Toeplitz matrices discussed in
~\cite[Example~6.5.4, Theorem~12.4.1]{lancaster1985}.

Let $X = VD$ and $B = X J X^{-1}$. Then
\begin{align}
B & = XJ X^{-1}\\
&= VD J D^{-1} V^{-1}\\
&= V D D^{-1} J V^{-1}\\
& = V J V^{-1}\\
& = A.
\end{align}
Therefore, both $V$ and $X$ are eigenvector matrices of $A$.
\end{IEEEproof}

In the following, it is assumed that $V$ satisfies Theorem~\ref{thm:norm1Jordanchain}. Theorem~\ref{thm:TV_ub} presents an upper bound of~\eqref{eq:TV_beforeineq}.
\begin{theorem}
\label{thm:TV_ub}
Consider matrix $A$ with $k$ distinct eigenvalues and $N\times r_{ij}$ matrices~$V_{ij}$  with columns comprising the $j$th Jordan chain of $\lambda_i$, $i=1,\dots,k$, $j=1,\dots,g_i$. Then the graph total variation $\mathrm{TV}_G(V_{ij})\leq \left|1-\lambda_i \right|+1$.
\end{theorem}
\begin{IEEEproof}
Let $\left\|V_{ij}\right\|_1 = 1$ and rewrite~\eqref{eq:TV_beforeineq}:
\begin{align}
\mathrm{TV}_G\left(V_{ij}\right)&\leq \left\|V_{ij}\right\|_1 \left\|I_{r_{ij}} - J_{ij} \right\|_1\\
&= \left\|I_{r_{ij}} - J_{ij} \right\|_1\label{eq:TV_ineq}\\
&= \left|1 - \lambda_i \right| +1.\label{eq:TV_ineqeval}
\end{align}
\end{IEEEproof}

Equations~\eqref{eq:TV_beforeineq},~\eqref{eq:TV_jordanchainvecs}, and~\eqref{eq:TV_ineqeval} characterize the (graph) total variation of a Jordan chain by quantifying the change in a set of vectors that spans the Jordan subspace~$\mathscr J_{ij}$ when they are transformed by the graph shift~$A$. While this total variation bound may not capture the true total variation of a spectral component, it can be generalized as an upper bound for all spectral components associated with a Jordan equivalence class. This concept is explored further in~\cite{deriEquiv2016}.
\begin{figure*}[hbt]
\centering
\begin{subfigure}[b]{.5\textwidth}
	\centering
  \includegraphics[width=1\linewidth]{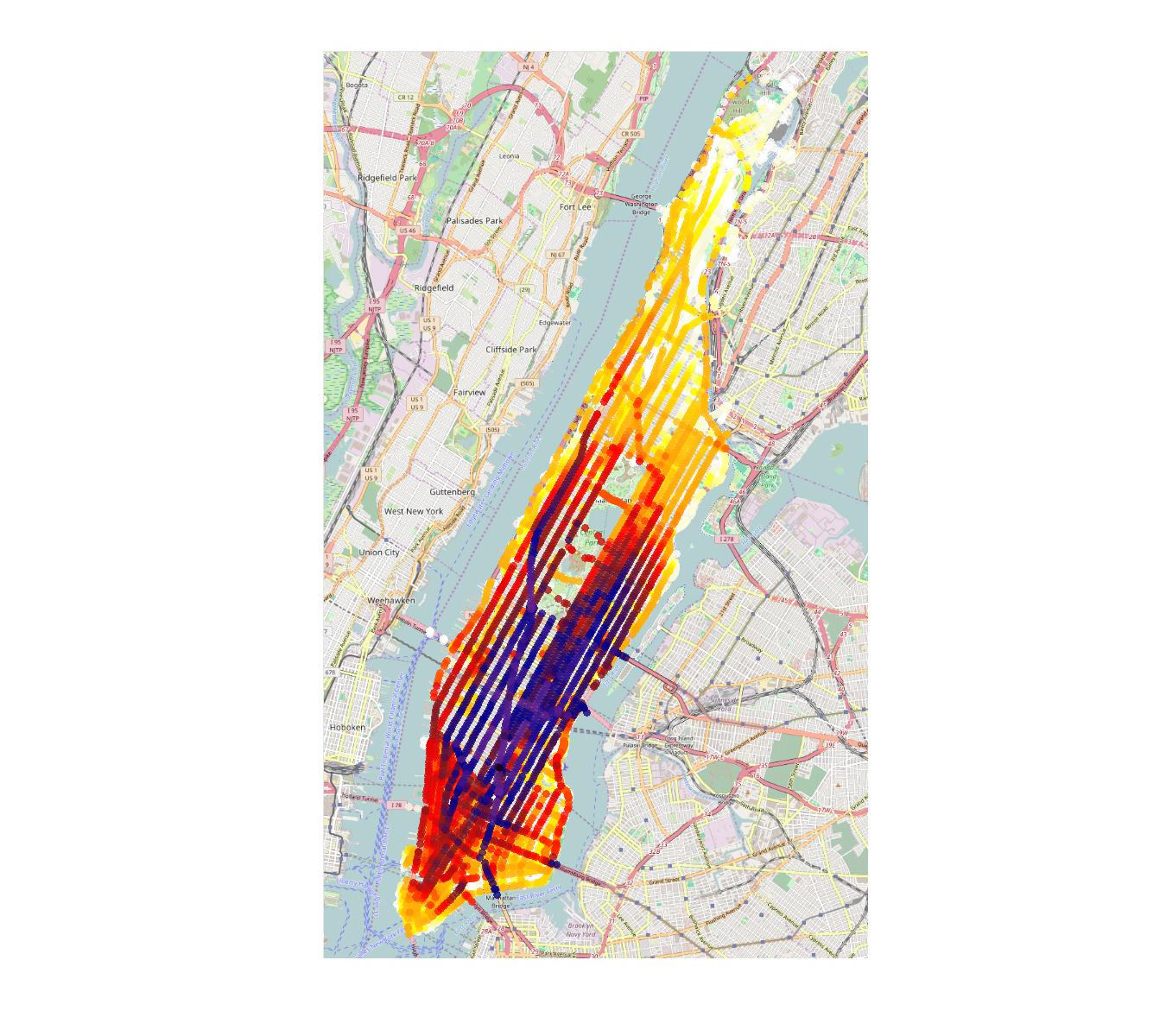}
  \caption{Average number of June-August trips, Fridays 9pm-10pm}
  \label{fig:sig}
\end{subfigure}%
\begin{subfigure}[b]{.5\textwidth}
	\centering
	\includegraphics[width=1\linewidth]{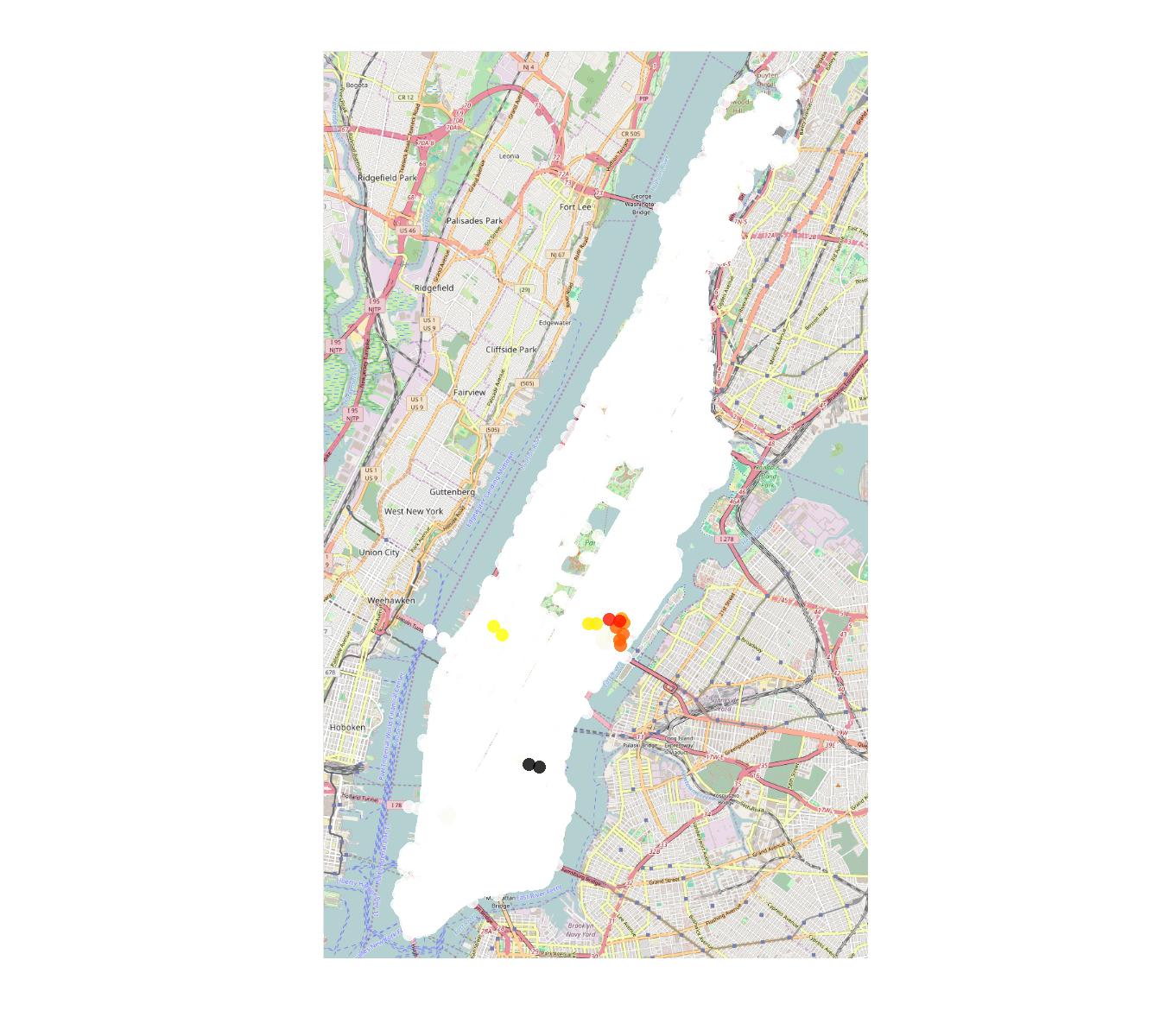}
	\caption{Maximum expressed eigenvector}
	\label{fig:eigenvec}
\end{subfigure}
\caption{Graph signal based on NYC taxi data (a) and the maximum expressed eigenvector (b).   (a) Colors denote log$_{10}$ bins of the four-year average number of trips for Fridays 9pm-10pm (699 log bins; white: 0--20, beige to yellow: 20--230, orange: 230--400, red: 400--550, blue: 550--610, purple to black: 610--2,700. (b) All colors except white indicate locations with a concentrated number of taxi trips. Black indicates the locations of maximum expression in the eigenvector. The plots were generated with ggmap~\cite{tool:ggmap} and OpenStreetMap~\cite{data:openstreetmap}.}
\label{fig:applic}
\end{figure*} 
\section{Application}
\label{sec:application}
We apply the graph Fourier transform~\eqref{eq:gft} to a signal based on 2010-2013 New York City taxi data~\cite{data:taxidata} over the Manhattan road network. The signal represents the four-year average number of trips that pass through each node of the road network as determined by Dijkstra path estimates from the start- and end-coordinates (latitude and longitude) provided by the raw data; the computation behind these estimates is described in~\cite{deriFranzMoura2016dijkstra}.  The road network consists of 6,408 nodes that represent latitude and longitude coordinates from~\cite{data:roadnetwork} that are  connected by 14,418 directed edges that represent one-way directions as verified by Google Maps~\cite{data:googlemapsmanhattan}. The adjacency matrix of the road network is defective with 253 Jordan chains of length~2 and 193 eigenvectors without Jordan chain vectors corresponding to eigenvalue zero (446 Jordan chains total). Details for the eigendecomposition of this matrix are described in~\cite{deriasilomar2015,deriAIM2016}.

Figure~\ref{fig:sig} shows the signal, consisting of the four-year June-August average number of trips at each node of the Manhattan road network for Fridays 9pm-10pm. Applying the GFT~\eqref{eq:gft} and computing the energies of the signal projections as described in Section~\ref{sec:gft:parseval} yields a highly expressed eigenvector shown in Figure~\ref{fig:eigenvec}. This eigenvector shows that most of the signal energy is concentrated at Gramercy Park (shown in black), north of Gramercy Park, and in Hell's Kitchen on the west side of Manhattan.
\section{Limitations}
\label{sec:limitations}
The graph Fourier transform presented in this paper solves the problem of uniqueness for defective adjacency matrices by projecting a signal onto Jordan subspaces instead of eigenvectors and generalized eigenvectors. This method relies on Jordan chain computations, however, which is sensitive to numerical errors and can be expensive when the number of chains to compute and the graph dimension are large and the computing infrastructure is memory-bound.  The computation can be accelerated by using equivalence classes over graph topologies, as we discuss in~\cite{deriEquiv2016}. These classes allow matching between a given graph to graphs of simpler topologies such that the graph Fourier transform of a signal with respect to these graphs is preserver. Since the equivalence classes may not be applicable to arbitrary graph structures, we explore inexact eigendecomposition methods in~\cite{deriAIM2016} to reduce the execution time and bypass the numerically unstable Jordan chain computation.

%
%
%
%
\section{Conclusion}
\label{sec:conclusion}
The graph Fourier transform proposed here provides a unique and non-ambiguous spectral decomposition for signals over graphs with defective, or non-diagonalizable, adjacency matrices. The transform is based on spectral projections of signals onto the Jordan subspaces of the graph adjacency matrix. This coordinate free graph Fourier transform is unique and leads to a unique spectral decomposition of graph signals. This paper shows that the signal projections onto the Jordan subspaces can be ranked by energy via a generalized Parseval's identity. Lastly, a total variation-based ordering of the Jordan subspaces is proposed. This allows ordering frequencies, and to define low-, high-, and band-pass graph signals.
\appendices
\section{Background on Jordan Decompositions}
\label{app:JordanDecomp}
\textbf{Direct sum.} Let $X_1,\dots,X_k$ be subspaces of vector space~$X$ such that~$X = X_1 + \cdots  + X_k$.  If~$X_i \cap X_j = \emptyset$ for all~$i\neq j$, then~$X$ is the \emph{direct sum} of subspaces $\{X_i\}_{i=1}^k$, denoted as~$X = \bigoplus_{i=1}^k X_i$. Any $x\in X$ can be written uniquely as~$x = x_1 +\cdots + x_k$, where $x_i\in X_i$,~$i=1,\dots,k$.

\textbf{Eigenvalues and multiplicities.}
Consider matrix~$A\in \mathbb C^{N\times N}$ with~$k$ distinct eigenvalues~$\lambda_1,\dots,\lambda_k$, $k\leq N$. 
The eigenvalues of~$A$ are the roots of the \emph{characteristic polynomial} $\varphi_A(\lambda) = \det(A - \lambda I) = \prod_{i=1}^{k}\left(\lambda - \lambda_i\right)^{a_i}$,  $I$ is the identity matrix, and exponent~$a_i$ represents the \emph{algebraic multiplicity} of eigenvalue $\lambda_i$, $i=1,\dots,k$. Denote by $\mathrm{Ker}(A)$ the kernel or null space of matrix $A$, i.e., the span of vectors $v$ satisfying $Av=0$. The \emph{geometric multiplicity}~$g_i$ of eigenvalue $\lambda_i$ equals the dimension of  null space  $\mathrm{Ker} \left(A-\lambda_i I\right)$. The \emph{minimal polynomial}~$m_A(\lambda)$ of~$A$ has form $m_A(\lambda) = \prod_{i=1}^{k} \left(\lambda - \lambda_i\right)^{m_i}$, where $m_i$ is the \emph{index} of eigenvalue~$\lambda_i$. The index~$m_i$ represents the maximum Jordan chain length or Jordan subspace dimension, which is discussed in more detail below.

\textbf{Generalized eigenspaces.} The eigenvectors and generalized eigenvectors of matrix $A\in\mathbb C^{N\times N}$ partition $\mathbb C^N$ into subspaces, some of which are spans of eigenvectors, eigenspaces, or generalized eigenspaces.
Subspace $\mathscr G_i = \mathrm{Ker} \left(A - \lambda_i I\right)^{m_i}$ is the \emph{generalized eigenspace} or \emph{root subspace} of~$\lambda_i$. The generalized eigenspaces are~$A$-invariant; that is, for all~$x\in \mathscr G_i$,~$Ax\in \mathscr G_i$. The subspace $\mathscr S_{ip} = \mathrm{Ker} \left(A - \lambda_i I\right)^p$, $p=0,1,\dots,N$, is the \emph{generalized eigenspace of order~$p$} for~$\lambda_i$. For~$p\geq m_i$,~$\mathscr S_{ip} = \mathscr G_i$.  The \textit{proper} eigenvectors~$v$ of $\lambda_i$, or simply eigenvectors of $\lambda_i$, are linearly independent vectors in~$\mathscr S_{i1}=\mathrm{Ker} \left(A - \lambda_i I\right)$, the \emph{eigenspace} of~$\lambda_i$. There are $g_i=\mathrm{dim}\,\mathscr S_{i1}=\mathrm{dim}\,\mathrm{Ker} \left(A - \lambda_i I\right)$ eigenvectors of $\lambda_i$. Subspaces~$\mathscr S_{ip}$ form a \emph{(maximal) chain} of $\mathcal G_i$ as depicted in Figure~\ref{fig:jordan_explain}; that is,
\begin{align*}
\{0\}\! = \!\mathscr S_{i0} \subset \mathscr S_{i1} \subset \! \cdots \subset \!\!\mathscr S_{i,m_i} = \mathscr S_{i,m_i + 1} = \cdots \subset \mathbb C^N
\end{align*}
where $m_i$ is the index of~$\lambda_i$. Vectors $v\in \mathscr S_{ip}$ but $v\notin \mathscr S_{i,p-1}$ are \emph{generalized eigenvectors of order~$p$} for~$\lambda_i$.

\textbf{Properties of Jordan subspaces.} A Jordan subspace~\eqref{eq:jordansubspace} is~$A$-invariant; that is, for all $x\in\mathscr J$,~$Ax\in\mathscr J$.
 The Jordan subspace~$\mathscr J$ is also \emph{cyclic} since it can be written by~\eqref{eq:jordanrecurse_subspace} as
\begin{equation}
\label{eq:jordancyclicspace}
\mathscr J = \mathrm{span}\left(v_{r}, (A- \lambda I)v_{r},\dots,(A-\lambda I)^{r-1}v_{r}\right)
\end{equation}
for $v_{r}\in \mathrm{Ker} (A-\lambda I)^{r}$, $v_r\neq 0$.

The number of Jordan subspaces corresponding to $\lambda_i$ equals the geometric multiplicity $g_i=\mathrm{dim}\,\mathrm{Ker} (A-\lambda I)$, since there are $g_i$ eigenvectors of $\lambda_i$. It can be shown that the Jordan spaces $\left\{\mathscr J_{ij}\right\}$, $j=1,\cdots, g_i$ and $i=1,\cdots,k$, are disjoint.

\textbf{Jordan decomposition.} Let~$V_{ij}$ denote the $N\times r_{ij}$ matrix whose columns form a Jordan chain of eigenvalue $\lambda_i$ of~$A$ that spans Jordan subspace~$\mathscr J_{ij}$.  Then the generalized eigenvector matrix~$V$ of $A$ is
\begin{equation}
\label{eq:V}
V = \begin{bmatrix}
V_{11}  \cdots  V_{1g_1} & \cdots &V_{k1} \cdots V_{kg_k}
\end{bmatrix},
\end{equation}
where $k$ is the number of distinct eigenvalues. The columns of $V$ are a \emph{Jordan basis}  of $\mathbb C^N$. Then~$A$ has block-diagonal \emph{Jordan normal form}~$J$ consisting of Jordan blocks
\begin{equation}
\label{eq:jordanblock}
\setlength\arraycolsep{12pt}
J(\lambda) = \begin{bmatrix}
\lambda&1& \\
&\lambda&\ddots     \\
&&\ddots &1\\
&  && \lambda
\end{bmatrix}.
\end{equation}
of size~$r_{ij}$; see, for example,~\cite{lancaster1985} or~\cite[p.125]{jordan1870}.  The Jordan normal form $J$ of $A$ is unique up to a permutation of the Jordan blocks. The \emph{Jordan decomposition} of~$A$ is $A = VJV^{-1}$.

\bibliographystyle{IEEEbib}
\bibliography{refs} 

\end{document}